\documentclass[runningheads]{llncs}
\usepackage[T1]{fontenc}
\usepackage{graphicx}
\usepackage{url}

\usepackage{amsmath,amssymb,amsfonts}
\usepackage{algorithm}
\usepackage{algorithmic}
\usepackage{subcaption}
\usepackage{graphicx}
\usepackage{textcomp}
\usepackage{stmaryrd}
\usepackage{xcolor}
\usepackage{multirow}
\usepackage[most]{tcolorbox}
\usepackage{colortbl}
\usepackage{pifont}
\usepackage{float}              
\usepackage{tabularx}
\usepackage{braket}
\usepackage{hyperref}
\usepackage{caption}
\usepackage{subcaption}
\usepackage{makecell}
\usepackage{multirow}
\usepackage{enumitem}
\setlist[itemize]{left=0pt}
\usepackage{subcaption}
\usepackage{listings}
\usepackage{quantikz}
\usepackage{tikz}
\usepackage{booktabs}
\usepackage{adjustbox}
\usepackage[table]{xcolor}
\usepackage{multirow}
\usepackage{makecell}
\usepackage{array} 
\renewcommand{\arraystretch}{1.2} 
\usepackage{placeins}

\usepackage{subcaption}
\usepackage{listings,xcolor,array}
\lstdefinestyle{qcode}{
  basicstyle=\ttfamily\scriptsize,
  numbers=left,
  numbersep=6pt,
  xleftmargin=1.2em,
  aboveskip=0pt, belowskip=0pt,
  linewidth=\linewidth,     
  columns=flexible,         
  keepspaces=true,
  breaklines=true,          
  breakatwhitespace=false,  
  breakindent=0pt,
  postbreak=\mbox{\textcolor{gray}{\tiny$\hookrightarrow$}\,} 
}
\newlength{\rowA}
\newlength{\rowB}

\lstdefinestyle{tightcodeSmall}{
  basicstyle=\ttfamily\scriptsize, 
  numbers=left, numbersep=4pt, numberstyle=\tiny,
  xleftmargin=1em,                 
  frame=single, framesep=1.5pt, rulecolor=\color{black!60},
  aboveskip=0pt, belowskip=0pt,    
  linewidth=\linewidth,            
  columns=flexible, keepspaces=true,
  breaklines=true, breakatwhitespace=false, breakindent=0pt,
  postbreak=\mbox{\textcolor{gray}{\tiny$\hookrightarrow$}\,} 
}

\usepackage[table]{xcolor}   
\usepackage{booktabs, tabularx, makecell, array}
\newcolumntype{Y}{>{\raggedright\arraybackslash}X}

\tcbset{
  colback=blue!5!white,    
  colframe=blue!40!black,  
  boxrule=0.5pt,           
  arc=2mm,                 
  left=1mm, right=1mm, top=1mm, bottom=1mm,  
  enhanced
}

\begin{document}
\title{QEMI: A Quantum Software Stacks Testing Framework via Equivalence Modulo Inputs}

\titlerunning{Quantum Equivalence Modulo Inputs}

%
%

\author{Junjie Luo\inst{1}\orcidID{0009-0008-7821-6879} \and
Shangzhou Xia\inst{2}\orcidID{0009-0006-2775-9633}  \and
Fuyuan Zhang\inst{3}\thanks{Fuyuan Zhang is with the State Key Laboratory of Blockchain and Data Security, Zhejiang University, Hangzhou, China}\orcidID{0009-0001-6560-5102} \and
Jianjun Zhao\inst{4}\orcidID{0000-0001-8083-4352}}

\authorrunning{J. Luo et al.}

%

\institute{Kyushu University, Japan \email{luo.junjie.609@s.kyushu-u.ac.jp} \and
Kyushu University, Japan \email{xia.shangzhou.218@s.kyushu-u.ac.jp} \and
Zhejiang University, China
\email{fuyuanzhang@zju.edu.cn} \and
Kyushu University, Japan
\email{zhao@ait.kyushu-u.ac.jp}}

\maketitle              
\begin{abstract}
As quantum algorithms and hardware continue to evolve, ensuring the correctness of the quantum software stack (QSS) has become increasingly important. However, testing QSSes remains challenging due to the oracle problem, i.e., the lack of a reliable ground truth for expected program behavior. Existing metamorphic testing approaches often rely on equivalent circuit transformations, backend modifications, or parameter tuning to address this issue. In this work, inspired by Equivalence Modulo Inputs (EMI), we propose Quantum EMI (QEMI), a new testing approach for QSSes. Our key contributions include: (1) a random quantum program generator that produces code with dead code based on quantum control-flow structures, and (2) an adaptation of the EMI technique from classical compiler testing to generate variants by removing dead code. By comparing the behavior of these variants, we can detect potential bugs in QSS implementations. We applied QEMI to Qiskit, Q\#, and Cirq, and successfully identified $11$ crash bugs and $1$ behavioral inconsistency. QEMI expands the limited set of testing techniques available for quantum software stacks by going beyond structural transformations and incorporating semantics-preserving ones into quantum program analysis. 

\keywords{Quantum computing \and Metamorphic testing \and Software testing \and Equivalent program variants}
\end{abstract}
\section{Introduction}
\label{sec:introduction}

Quantum computing promises advantages on specific problem classes~\cite{grover1996,Shor_1997}. Superposition and entanglement, as key qubit properties, enable algorithmic speedups but also make it hard for classical languages to faithfully represent and manipulate quantum states.

Quantum software stacks (QSSes) address this gap by providing the tools to construct, compile, and execute quantum programs on simulators or hardware. They offer high-level abstractions for expressing quantum logic while handling backend-specific details. A typical QSS comprises a domain-specific interface for building circuits, utilities for circuit transformation/optimization, and runtime/backends for execution. Representative systems include \textit{Qiskit} (IBM)~\cite{qiskit}, \textit{Q\#} (Microsoft)~\cite{qsharp}, and \textit{Cirq} (Google)~\cite{Cirq}, each with its own language syntax, runtime semantics, and toolchain.




Unlike classical software systems, quantum software often integrates both quantum and classical components. For instance, a quantum algorithm may include classical control logic (\textit{e.g.}, conditionals or loops around quantum operations) or quantum control logic (\textit{e.g.}, using qubits as control conditions). This hybrid nature, combined with the probabilistic behavior of quantum programs, presents unique challenges for the design and verification of quantum software systems. As these stacks continue to evolve, ensuring their correctness and reliability has become an important concern in the development of practical quantum software.


Testing plays a crucial role in validating the correctness of quantum software stacks (QSSes). QSS testing faces two major challenges. As mentioned in~\cite{morphq}, QSS testing now faces two challenges: the lack of quantum programs for testing and the oracle problem. Existing approaches often leverage random quantum program generators in combination with techniques such as differential and metamorphic testing. These strategies have proven effective at identifying real-world bugs in QSSes. For instance, \textit{QuteFuzz}~\cite{qutefuzz}, which applies differential testing on randomly generated programs, uncovered 17 bugs across multiple versions of Qiskit, Cirq, and Pytket~\cite{pytket}. Other tools such as \textit{QDiff}~\cite{qdiff} and \textit{MorphQ}~\cite{morphq} also demonstrate strong effectiveness, detecting 4 and 13 crash bugs, respectively.

However, existing methods still have room for improvement. Current random program generators largely emphasize composing quantum gate operations while paying limited attention to broader QSS APIs (e.g., the predefined circuits in \textit{qiskit.circuit.library}), which risks overlooking API-specific bugs during testing. In addition, many existing techniques prioritize circuit-level or transformation-based equivalence, placing less emphasis on language-level semantic equivalence between quantum programs. To address these gaps, this paper revisits \textit{Equivalence Modulo Inputs} (EMI)~\cite{emi}—a technique that generates program variants which are semantically equivalent with respect to a given input set by modifying code that does not affect behavior for that input—so as to enable differential detection of implementation bugs. Applying EMI to quantum software stacks, however, poses unique challenges: due to superposition-induced probabilistic outputs, even with deterministic quantum inputs, the program behavior can be non-deterministic, making input-irrelevant code hard to identify. Moreover, \textcolor{black}{some} QSSes execute quantum and classical code separately, limiting the use of classical variables to locate dead code within quantum control structures.

In our approach, we propose \textbf{QEMI}\footnote{We provide access to QEMI artifacts at \textcolor{blue}{\url{https://github.com/Xzore19/QEMI}}}, a framework that adapts EMI to quantum software stacks. QEMI targets quantum-specific control-flow structures and generates semantically equivalent variants by removing dead code inside quantum conditional statements. Unlike classical EMI, where dynamic execution can reveal dead code, the probabilistic nature of quantum outputs makes such a strategy impractical. Instead, QEMI analyzes quantum control statements and, by preparing specific quantum states in advance, ensures that designated code blocks do not affect program outputs. Based on this insight, QEMI defines a set of dead-code patterns and constructs the corresponding quantum states during program generation. \textcolor{black}{We focus on dead-code insertion, and richer equivalence-preserving rewrites inside these regions are left as future work.} Using this static approach, QEMI uncovered 12 bugs across three widely used QSSes (Qiskit, Q\#, and Cirq). To improve efficiency, QEMI further incorporates an early-stop strategy that terminates once a predefined statistical-confidence threshold is met for distinguishing output distributions, thereby reducing the number of final measurements per execution and yielding at least a $53.83\%$ speedup in our experiments.

The main contributions of this work include: 
\begin{itemize}[left=0pt]
    \item Developing a quantum random program generator capable of producing programs with both classical and quantum control flow, designed for testing with the latest APIs.
    
    \item Integrating a static pattern-based EMI approach for quantum programs into the random program generator.

    \item Designing an early stopping strategy for detecting consistency between measurement distributions during quantum program comparison.

    \item Implementing the framework on popular QSSes, including Qiskit, Q\#, and Cirq, and uncovering $12$ unique real-world bugs, including both distributional and crash-related errors.
\end{itemize}

The rest of the paper is organized as follows. Section~\ref{sec:background} provides background information relevant to our work. Section~\ref{sec:methodology} describes the design of QEMI. Section~\ref{sec:result} presents our experimental setup and results. Threats to validity are discussed in Section~\ref{sec:threats}. We review related work in Section~\ref{sec:related work}, and conclude the paper in Section~\ref{sec:conclusion}.

\section{Background}
\label{sec:background}

\subsection{Quantum Computing}

Quantum computing provides a fundamentally new approach to information processing by leveraging quantum mechanical phenomena, including superposition, entanglement, and interference. Unlike classical computation, which is based on deterministic operations over binary bits, quantum computation is built upon quantum circuits operating on qubits.

\paragraph{Qubits and Quantum Gates.}
The basic unit of quantum information is the \textit{qubit}. A qubit can exist in a superposition of classical states 0 and 1, and its state is generally expressed as:
\[
\ket{\psi} = \alpha \ket{0} + \beta \ket{1},
\]
where $\alpha$ and $\beta$ are complex numbers that satisfy the normalization condition $|\alpha|^2 + |\beta|^2 = 1$. To manipulate qubit states, quantum gates are applied. These gates are unitary matrices that transform the state of one or more qubits in a reversible manner. Common quantum gates include Pauli-X, Hadamard, and phase gates for single-qubit operations, as well as multi-qubit gates like CNOT, Toffoli, and SWAP. By composing these gates, one can construct quantum circuits capable of encoding complex quantum algorithms.

\paragraph{Quantum Circuits and Measurement.}
A quantum program is typically implemented as a circuit that applies a sequence of quantum gates to an initial state. If the system starts in state $\ket{\psi_0}$ and performs operations $U_1, U_2, \dots, U_k$, then the final state is as follows:
\[
\ket{\psi_f} = U_k U_{k-1} \dots U_1 \ket{\psi_0}.
\]
After the circuit is executed, a measurement projects the final quantum state onto a classical bitstring. Because measurement outcomes are probabilistic, a quantum program is usually executed multiple times to estimate the output distribution. These repeated executions are called \emph{shots} (one shot produces one bitstring sample). The statistical nature of measurement plays a crucial role in many quantum algorithms, where desired results appear with high probability only after repeated trials.

\subsection{Equivalence Modulo Inputs}

Equivalence Modulo Inputs (EMI) is a relaxed notion of semantic equivalence, originally formalized by Le et al.~\cite{emi}. Let $\mathcal{L}$ be a deterministic programming language with semantics $\llbracket \cdot \rrbracket$.

Two programs $P, Q \in \mathcal{L}$ are said to be \emph{EMI-equivalent} with respect to an input set $I \subseteq \mathrm{dom}(P) \cap \mathrm{dom}(Q)$ if and only if:
\[
\forall i \in I.\ \llbracket P \rrbracket(i) = \llbracket Q \rrbracket(i)
\]
This relation is denoted by $\llbracket P \rrbracket =_I \llbracket Q \rrbracket$, indicating that $P$ and $Q$ exhibit identical behavior for all inputs in $I$, even if they differ syntactically or produce different outputs on other inputs.

Given a program $P$ and an input set $I \subseteq \mathrm{dom}(P)$, the set of all programs $Q$ that are EMI-equivalent to $P$ with respect to $I$ constitutes the \textbf{EMI variants} of $P$:
\[
\{ Q \in \mathcal{L} \mid \llbracket P \rrbracket =_I \llbracket Q \rrbracket \}
\]

When $I = \{\texttt{void}\}$, meaning the programs do not take inputs, EMI reduces to classical semantic equivalence. In this case:
\[
\llbracket P \rrbracket = \llbracket Q \rrbracket \quad \Rightarrow \quad \llbracket P \rrbracket =_I \llbracket Q \rrbracket
\]
If a compiler produces different outputs for $P$ and $Q$ under a fixed execution configuration, this indicates a potential semantic bug. EMI leverages this principle by constructing program variants that are guaranteed to behave identically under the same configuration. This enables compiler testing without the need for a reference implementation or ground truth, while preserving the semantic validity of test programs.

\subsection{Control Flow in Quantum Programs}
\label{ssec:ctlflow}

Since control flow in quantum programs differs in certain ways from that in classical programs, this section provides a detailed overview of control flow constructs in quantum programming. 

\subsubsection{Quantum Conditions} 

\begin{figure}[htbp]
  \centering
  \small
  \begin{subfigure}[b]{0.4\textwidth}
  \centering
    \begin{quantikz}[row sep={1cm,between origins}]
\lstick{$q_0$} & \ctrl{1} & \qw \\
\lstick{$q_1$} & \targ{}  & \qw
\end{quantikz}
    \caption{Quantum control gate.}
    \label{fig:control gate}
  \end{subfigure}
  \begin{subfigure}[b]{0.4\textwidth}
    \includegraphics[width=\linewidth]{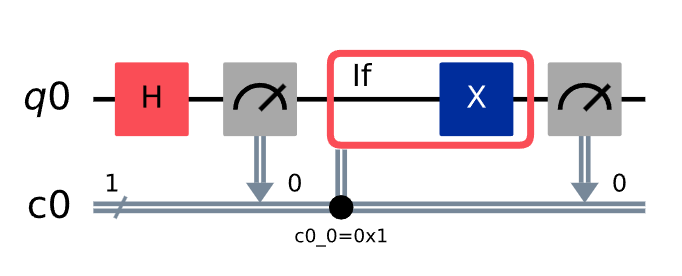}
    \caption{Quantum dynamic circuit.}
    \label{fig:dynamic circuit}
  \end{subfigure}
  \caption{Two types of quantum conditions.}
  \label{fig:code_comparison}
\end{figure}

To clearly explain the control flow in quantum programs, we begin by introducing the concept of a quantum condition. In classical programs, a condition typically refers to a Boolean expression whose evaluation determines whether a certain block of code is executed or repeated. This expression involves only classical variables, and thus its expressive power is limited in the context of quantum programs. Unlike classical conditions that rely exclusively on classical variables, quantum conditions refer to control constructs that depend on quantum states (Figure~\ref{fig:control gate}) or on qubit measurement outcomes (Figure~\ref{fig:dynamic circuit}), including direct use of such outcomes as control inputs. It allows quantum programs to express more sophisticated control flow structures. 

\subsubsection{Quantum Control Flow.}

For this work, any control flow that involves quantum operations in either the condition or in the controlled operation is considered a quantum control flow. QEMI focuses on quantum-target control flows. 
Based on this focus, we identify several common control flow structures in quantum programs that serve as the basis for subsequent random program generation and the design of dead code patterns. 

\paragraph{Conditional Branching Structure.} Quantum programs mainly include three forms of conditional branching structures.

The first is to use a classical condition to control quantum operations, where the overall structure closely resembles conditional code constructs in classical programs. When the classical condition evaluates to false, certain regions of the program become dead code. For example, using the \textit{if\_else} statement implements the logic:
\begin{lstlisting}[basicstyle=\ttfamily\small,
        columns=fullflexible,
        % numbers=left,
        % frame=single, 
        xleftmargin=2em,]
    if (a == 1) { Op(tgt) }
\end{lstlisting}
where \textit{a} is a classical variable and \textit{Op(tgt)} is a quantum operation, and when $a \neq 1$, the \textit{Op(tgt)} will not execute.

The second is an extension of the classical \textit{if-else} statement to the quantum setting, where quantum variables are used as a condition. In this case, when the quantum variable does not match a specific quantum state, which is represented as an int value, the code within the true branch will not be executed. For example, in Q\#, the function \textit{ApplyIfEqualL(Op, c, ctrl, tgt)}~\cite{qsharp-api} implements the logic: 
\begin{lstlisting}[basicstyle=\ttfamily\small,
        columns=fullflexible,
        % numbers=left,
        % frame=single, 
        xleftmargin=2em,]
    if (c == ctrl) { Op(tgt) }
\end{lstlisting}

where the operation \textit{Op} is applied to the target quantum register \textit{tgt} if the quantum register \textit{ctrl} equals the specific quantum state c, which is represented as a classical value \textit{c}.
Since \textit{ctrl} could be in superposition states, the control flow governed by such statements will only satisfy the non-execution property of certain code blocks for a very limited subset of quantum states.


The third is a quantum-native control structure that is integrated into quantum operations as a fundamental concept of quantum computing. Certain quantum operations can be directly controlled by specific qubit states. For example, the controlled-NOT gate ($CX$) is typically represented by $CX\ket{ctrl,tgt}=\ket{ctrl,tgt\oplus ctrl}$.
From a semantic point of view, the $CX$ gate can also be interpreted as a form of conditional branching: when the control qubit \textit{ctrl} is in the state \textit{1}, apply the \textit{X} gate to the target qubit \textit{tgt}:

\begin{lstlisting}[basicstyle=\ttfamily\small,
        columns=fullflexible,
        % numbers=left,
        % frame=single, 
        xleftmargin=2em,]
    if (ctrl == 1) { X(tgt) } 
\end{lstlisting}

\paragraph{Loop Structure.} 

Loop-related constructs in quantum programs are similar to those in classical programs and mainly include iteration and conditional loops. Some commonly used loop constructs include:
\begin{itemize}
  \item \textbf{Iterations}, such as: \textit{for i in 1..n \{ Operation \}}
  \item \textbf{Condition-controlled loops}, including:
  \begin{itemize}
    \item \textit{while condition \{ Operation \}}
    \item \textit{repeat \{ Operation \} until condition fixup \{ Operation \}}
  \end{itemize}
\end{itemize}

The \textit{condition} in these statements can be implemented using either classical variables or quantum states and measurement results. The fundamental logic of quantum conditions is essentially the same as that in conditional branching structures. \textcolor{black}{While a quantum condition can be logically represented as a classical statement, during execution, the inherent probabilistic nature of quantum measurements leads the \textit{condition} to evaluate to \textit{True} or \textit{False} in a probabilistic manner.}



\section{Methodology}
\label{sec:methodology}

\begin{figure}
    \centering
    \includegraphics[width=1\linewidth]{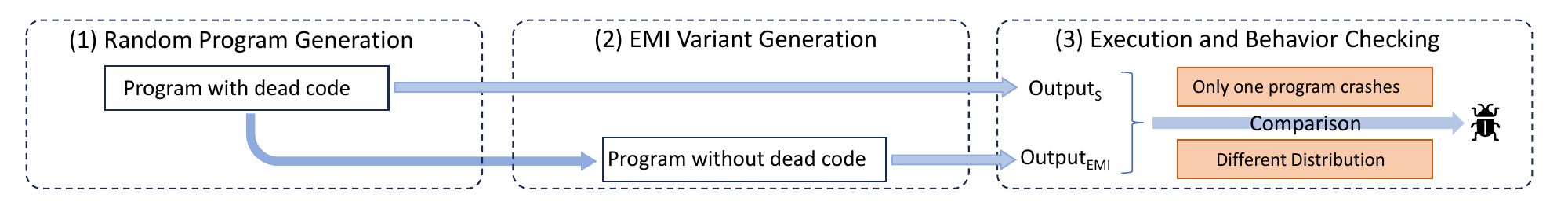}
    \caption{The workflow of QEMI. }
    \label{fig:workflow}
\end{figure}

Figure~\ref{fig:workflow} presents an overview of QEMI, a framework for testing QSSes based on EMI. QEMI consists of three main steps. First, a random program generator constructs a seed program by randomly selecting classical or quantum control structures, basic quantum gates, and API calls (Section~\ref{ssec:randqpgen}). During generation, dead code is inserted based on predefined patterns (Section~\ref{ssec:dcpattern}). Second, a static analysis pass removes the inserted dead code, yielding an EMI variant that is semantically equivalent to the original program (Section~\ref{ssec:qemivariantgen}). Finally, QEMI executes both the original and variant, checking for behavioral discrepancies such as crashes or divergences in the output distributions to identify potential bugs in QSS (Section~\ref{ssec:checking}).



\subsection{Dead Code Pattern Design}
\label{ssec:dcpattern}

Due to the presence of control flow structures, dead code may exist in quantum programs. However, analyzing the presence of dead code in randomly generated quantum programs presents unique challenges. In classical software testing, dead code is typically identified by static reachability analysis or dynamic execution profiling. However, both approaches are difficult to apply in the context of quantum programs.

First, reachability analysis remains underdeveloped at the level of high-level quantum programming languages. While progress has been made in reachability analysis at the circuit level (\textit{e.g.}, for quantum Markov chains~\cite{QReach}), there remains a lack of practical tools capable of analyzing control flow in high-level quantum programming languages. Developing such tools represents an important direction for future work. 

Second, classical dead-code identification based on deterministic inputs (e.g., dynamic coverage tools like Python’s \textit{coverage}) is incompatible with quantum control flow. Such tools execute a program on fixed inputs and mark lines as executed or not; in classical EMI, this reveals input-irrelevant code for generating semantically equivalent variants.
In quantum programs, however, behavior remains probabilistic even from a well-defined initial state. For example, $\ket{\phi}=0.8\ket{0}+0.6\ket{1}$ collapses to $\ket{1}$ with probability $0.64$ and to $\ket{0}$ with $0.36$, so a path that seems dead in one run may be taken in another. Hence, a single execution trace cannot reliably classify dead code and may yield false positives when comparing “equivalent” program variants.
This limitation is compounded by the structure of many frameworks (e.g., Qiskit), where quantum and classical code are constructed and handled separately.
Therefore, classical methods based on deterministic inputs cannot effectively identify dead code regions within quantum-specific control-flow structures.

To address these limitations, we adopt a static strategy: \textcolor{black}{instead of dynamically running programs and identifying dead code regions via coverage analysis, we generate corresponding dead-code patterns based on different types of quantum control statements.} These patterns ensure that the dead code regions remain unexecuted under any input set, thereby preserving semantic equivalence among program variants. Furthermore, by replacing dead code regions with alternative quantum operations, we can efficiently expand the number of program variants. In QEMI, we nest multiple dead code patterns to generate more complex control flow structures.

\begin{figure}[t]
  \centering

  \begin{subfigure}[t]{0.32\textwidth}\vspace{0pt}
    \begin{minipage}[t][\rowA][t]{\linewidth}\vspace{0pt}
\lstset{style=qcode}
\begin{lstlisting}
qc.measure(qreg, condition)
with qc.if_test(condition, 0) as else_0:
    # DEADCODE START
    qc.x(0)
    # DEADCODE END
with else_0:
    qc.h(0)
\end{lstlisting}
    \end{minipage}
    \caption{if\_test\_dead}\label{subfig:if_test}
  \end{subfigure}\hfill
  \begin{subfigure}[t]{0.32\textwidth}\vspace{0pt}
    \begin{minipage}[t][\rowA][t]{\linewidth}\vspace{0pt}
\lstset{style=qcode}
\begin{lstlisting}
qc.measure(qreg, condition)
with qc.while_loop(condition, 0):
    # DEADCODE START
    qc.x(0)
    # DEADCODE END
\end{lstlisting}
    \end{minipage}
    \caption{while\_dead}\label{subfig:while_dead}
  \end{subfigure}\hfill
  \begin{subfigure}[t]{0.32\textwidth}\vspace{0pt}
    \begin{minipage}[t][\rowA][t]{\linewidth}\vspace{0pt}
\lstset{style=qcode}
\begin{lstlisting}
qc.measure(qreg, condition)
with qc.switch(condition) as case:
    with case(0):
        # DEADCODE START
        qc.x(0)
        # DEADCODE END
    with case(1):
        qc.h(0)
\end{lstlisting}
    \end{minipage}
    \caption{\textit{switch\_dead}}\label{subfig:switch_dead}
  \end{subfigure}

  \vspace{0.4em} 

  \begin{subfigure}[t]{0.32\textwidth}\vspace{0pt}
    \begin{minipage}[t][\rowB][t]{\linewidth}\vspace{0pt}
\lstset{style=qcode}
\begin{lstlisting}
with qc.for_loop(range(0)) as i:
    # DEADCODE START
    qc.x(0)
    # DEADCODE END
\end{lstlisting}
    \end{minipage}
    \caption{for\_zero}\label{subfig:for_zero}
  \end{subfigure}\hfill
  \begin{subfigure}[t]{0.32\textwidth}\vspace{0pt}
    \begin{minipage}[t][\rowB][t]{\linewidth}\vspace{0pt}
\lstset{style=qcode}
\begin{lstlisting}
with qc.for_loop(range(5)) as i:
    qc.h(1)
    qc.continue_loop()
    # DEADCODE START
    qc.x(0)
    # DEADCODE END
\end{lstlisting}
    \end{minipage}
    \caption{for\_continue}\label{subfig:for_continue}
  \end{subfigure}\hfill
  \begin{subfigure}[t]{0.32\textwidth}\vspace{0pt}
    \begin{minipage}[t][\rowB][t]{\linewidth}\vspace{0pt}
\lstset{style=qcode}
\begin{lstlisting}
with qc.for_loop(range(5)) as i:
    qc.h(1)
    qc.break_loop()
    # DEADCODE START
    qc.x(0)
    # DEADCODE END
\end{lstlisting}
    \end{minipage}
    \caption{for\_break}\label{subfig:for_break}
  \end{subfigure}

  \vspace{-0em}
  \caption{Some examples for dead code patterns in Qiskit. For
  \textit{if\_test\_dead}, \textit{while\_dead}, and \textit{switch\_dead}, we first prepare $\ket{1}$ in \textit{qreg}. After measurement, the \textit{condition} equals 1 deterministically.}
  \label{fig:dcqkexample}
\end{figure}

\textcolor{black}{Figure~\ref{fig:dcqkexample} presents several dead code patterns in Qiskit. The dead code patterns that we designed are mainly divided into two categories:}

\begin{itemize}
    \item \textbf{Input-dependent:} These patterns ensure that the dead code region remains unexecuted only when a specific quantum state is provided as input. For example, in Figure~\ref{subfig:if_test}, if the quantum state does not match the target state $\ket{1}$, the measurement output (line 1) may yield 0, potentially triggering the execution of the dead code region. To handle these, we introduce an additional quantum register \textit{qreg} to prepare the desired target quantum state $\ket{1}$. 

    \item \textbf{Input-independent:} These patterns guarantee that the dead code region is never executed under any input. For example, in Figure~\ref{subfig:for_continue}, the semantics of \textit{qc.continue\_loop} ensure that the code following is never executed.
\end{itemize}



These patterns are deliberately constructed to ensure that the enclosed dead code is never executed under any condition, while preserving the syntactic and semantic structure of valid quantum control flows. Note that not all QSSes support every dead code pattern listed above; in practice, applicable patterns are selected based on the capabilities of each QSS.

\subsection{\textcolor{black}{Program Generation}}
\label{ssec:proggen}

We next describe how QEMI generates EMI variants by applying dead code patterns and how this mechanism is incorporated into the random program generation process. 

\subsubsection{Quantum EMI Variants generation}
\label{ssec:qemivariantgen}

Based on defined dead-code patterns, QEMI instantiates each pattern with concrete operations, such as quantum subcircuits, nested control-flow constructs, or nested dead code, to form a complete dead-code block. These operations are typically encapsulated into a callable function, which is then invoked at the designated dead code location. The inserted region is explicitly marked as semantically unreachable to ensure that it does not affect the program behavior. 
\textcolor{black}{While Qiskit allows users to construct reusable control logic through Python functions, Q\# exposes quantum control structures as built-in operations. These control-flow APIs encapsulate conditional logic at the language level, providing a modular, structured way to express complex control behavior. To demonstrate this feature, we provide a Q\# example. Figure~\ref{fig:dcsample} presents an example of a Q\# code segment generated from a dead code pattern and its corresponding EMI variant. In this case, the operation \textit{ApplyControlledOnInt(7, DeadBlock, ctrl, q)} is semantically equivalent to the conditional construct: }

\begin{lstlisting}[basicstyle=\ttfamily\small,
        columns=fullflexible,
        % numbers=left,
        % frame=single, 
        xleftmargin=2em,]
    if (ctrl == 7) { DeadBlock(q) }
\end{lstlisting} 
Since \textit{ctrl} is initialized to \textit{0}, the condition \textit{ctrl == 7} always evaluates false, which means that \textit{deadBlock (q)} is never executed. As a result, \textit{ApplyControlledOnInt(7, DeadBlock, ctrl, q)} does not affect the behavior of the program. During code generation, this region is marked as dead (e.g., lines~7 and~9 in Figure~\ref{subfig:dcsample}), and can be removed directly to generate an EMI variant (Figure~\ref{subfig:emivarsample}).

\begin{figure}[htb]
  \centering
  \begin{subfigure}[t]{0.49\columnwidth}\vspace{0pt}
\lstset{style=tightcodeSmall}
\begin{lstlisting}
operation DeadBlock(q : Qubit[]) : Unit is Adj + Ctl {
    Rzz(5.86706, q[1], q[2]);
}

operation ApplyRandomBlock1(q : Qubit[]) : Unit {
    use ctrl = Qubit[3];
    // --- DEADCODE START ---
    ApplyControlledOnInt(7, DeadBlock, ctrl, q);
    // --- DEADCODE END ---
}
\end{lstlisting}
    \subcaption{DCB in the original program.}
    \label{subfig:dcsample}
  \end{subfigure}\hfill
  \begin{subfigure}[t]{0.49\columnwidth}\vspace{0pt}
\lstset{style=tightcodeSmall}
\begin{lstlisting}
operation DeadBlock(q : Qubit[]) : Unit is Adj + Ctl {
    Rzz(5.86706, q[1], q[2]);
}

operation ApplyRandomBlock1(q : Qubit[]) : Unit {
    use ctrl = Qubit[3];




}
\end{lstlisting}
    \subcaption{DCB in QEMI variant.}
\label{subfig:emivarsample}
  \end{subfigure}

  \caption{An example for the dead code block(DCB) generation.}
  \label{fig:dcsample}
  \vspace{-0mm}
\end{figure}


\subsubsection{Random Quantum Program generation}
\label{ssec:randqpgen}

The quantum programs generated by QEMI mainly consist of the following four components:

\begin{itemize}
    \item \textbf{Basic Prologue Part: }
    To ensure that the quantum program runs correctly, the generator loads the required quantum libraries, the basic program configurations, and the chosen random seed into the program module.
    \item \textbf{Quantum Operation Part: }
    Based on the configuration settings, we initialize the required quantum and classical registers. For each quantum operation, the generator randomly selects from basic quantum gates, quantum API calls, and composite gates constructed from subcircuits, with the corresponding random parameters.
    
    \item \textbf{Dead Code Part: }
    Within the selected dead code patterns, the generator randomly inserts quantum operations. Due to the inherent uncertainty of quantum measurement outcomes, some dead code patterns (such as \textit{if\_test}, \textit{while\_loop}, \textit{ApplyIfEqual}, etc.) only exhibit strictly non-executable behavior under specific quantum states. For these patterns, we allocate additional quantum registers to prepare the desired quantum state, ensuring that the condition in the quantum control flow deterministically evaluates to either \textit{True} or \textit{False}. Given the transformable nature of dead code, the generator also outputs this module and marks the non-executable regions.
    \item \textbf{Optimization Pass Part (Optional): }
    Since some quantum platforms include built-in program optimization processes, we randomly apply optimization passes to the generated programs to ensure a more thorough testing of QSSes.
\end{itemize}

Based on the configuration settings, we construct the corresponding executor and parameter information for each generated quantum program.


\subsection{Execution and Behavior Checking}
\label{ssec:checking}

The \textit{Hellinger distance}~\cite{Hellinger1909} is a widely used metric to quantify the similarity between two discrete probability distributions. Given two distributions $P = (p_1, p_2, \dots, p_n)$ and $Q = (q_1, q_2, \dots, q_n)$ over the same finite domain, the Hellinger distance is defined as:
\begin{equation}
H(P, Q) = \frac{1}{\sqrt{2}} \sqrt{ \sum_{i=1}^{n} \left( \sqrt{p_i} - \sqrt{q_i} \right)^2 }
\label{eq:hellinger}
\end{equation}
The value of $H(P, Q)$ lies in the range $[0, 1]$, where $H(P, Q) = 0$ if and only if $P = Q$, and $H(P, Q) = 1$ when $P$ and $Q$ have a disjoint distribution. In our context, $P$ and $Q$ typically represent the normalized output distributions of the two quantum circuits being compared. We prefer the Hellinger distance over other divergences such as the Kolmogorov-Smirnov (K-S) test~\cite{KS1}, \cite{KS2} or cross-entropy~\cite{CE1}, ~\cite{CE2}, as it remains well defined on zero-probability events and aligns naturally with quantum fidelity measures~\cite{advofhellinger}. Although we recommend using the Hellinger distance as the primary evaluation metric, our method also supports alternative options such as the K-S test and cross-entropy, allowing users to choose according to their specific needs. In addition, users can extend QEMI by designing and integrating custom metrics into the framework.

\paragraph{Early Termination for Distribution Divergence Detection}

While we can assess the similarity between two distributions using the Hellinger distance to determine whether two quantum programs exhibit equivalent behavior, in practice, there is a trade-off between the confidence of the Hellinger estimate and the measurement cost. Existing research~\cite{lowboundofhd} shows that, for a Hellinger threshold $\delta$, determining whether two distributions are identical versus $\delta$-far in Hellinger distance with $2/3$ confidence requires at least

\begin{equation}
S_\text{std}=S(\delta, N) = \min\left\{ \frac{N^{2/3}}{\delta^{8/3}},\; \frac{N^{3/4}}{\delta^2} \right\} , 
\label{eq:hellinger-lower-bound}
\end{equation} 

where $N=2^n$ is the size of the output space for a quantum program that acts on $n$ qubits. For example, setting $\delta = 0.1$ and $n = 8$ results in a sample requirement of approximately $6,\!400$ measurements. 

However, in practical test scenarios, a confidence level of $2/3$ results in a relatively high error rate, potentially incurring additional cost due to the need for manual verification of samples flagged as inconsistent~\cite{morphq}. To raise the confidence level to a target of $1 - \eta$, it is necessary to apply confidence amplification~\cite{confidenceampl} by repeating the measurement process multiple times and aggregating the results. 
While confidence amplification boosts the reliability of the Hellinger estimate by repeating independent comparisons, it also leads to a proportional increase in the total number of measurements. Using too few measurements reduces the confidence of the estimated distance, while excessive measurements incur significant overhead. To reduce measurement cost without compromising statistical confidence, we aim to design an adaptive stopping strategy that terminates the comparison once sufficient evidence of equivalence has been observed.

For programs generated by QEMI, we have a useful prior: the two quantum programs produced by QEMI are expected to yield identical output distributions, which implies that $H(P, Q) = 0$. This implies that as the number of measurements increases, the Hellinger distance between the two resulting distributions should asymptotically converge to zero. 

Moreover, due to the inherent indeterminacy of quantum behavior, the output space size $2^n$ specified by a circuit only represents an upper bound. Causes that reduce the effective output space include entanglement between qubits, the use of phase gates that do not affect measurement probabilities, and, more directly, the absence of randomness-inducing operations on certain qubits. As a result, the actual set of outcomes that appear with non-negligible probability may be much smaller. 

Based on the above reasoning, we design an early-stopping strategy for QEMI when using the Hellinger distance to compare output distributions. For each randomly generated quantum program and its corresponding EMI variant, we repeatedly performed measurements with a limited number of shots in each round. Specifically, the number of shots per round is defined as
\begin{equation}
S_{\text{round}(\delta, N)} = S(\delta, N^{1/2}). 
\label{eq:earlystop-shot}
\end{equation}
This design leverages the two priors mentioned above. First, by using a smaller number of shots per round, we can observe the convergence of \( H(P, Q) \) to below the threshold \(\delta\) quickly, assuming that the programs are indeed equivalent. Second, the choice of using \( N^{1/2} \) as the effective output space size reflects our assumption that the true support of the output distribution is much smaller than the full \( N = 2^n \) space. By iteratively executing rounds with this reduced shot count, we aim to minimize the total number of measurements required to confidently determine whether \( H(P, Q) < \delta \).

After each measurement round, we compute the Hellinger distance between the two distributions. If the distance is smaller than~$\delta$, we perform one additional measurement round using the same number of shots $S_{\text{round}}$. If the Hellinger distance remains below~$\delta$ for two consecutive rounds, we terminate the process and conclude that there is no significant difference between the distributions. 

However, if the total number of measurements reaches the maximum allowed shot budget $S_{\text{max}}$ without observing two consecutive rounds with Hellinger distance below~$\delta$, we consider the distributions to be different. Here $S_\text{max}$ is a user-defined threshold: a larger value corresponds to a higher confidence level. In practice, we empirically set
\begin{equation}
    S_\text{max}=2\cdot S_\text{std}, 
\end{equation}

which demonstrates its effectiveness in practice. 

\subsection{Algorithm}

Algorithm~\ref{alg:alg1} describes how QEMI composes the three main steps. The first step of the entire process is to select the appropriate compiler configuration set (line 3), program generator (line 4), and quantum program executor (line 5) for the specified quantum programming platform. Based on the Hellinger threshold $\delta$ and the number of qubits $n$, QEMI calculates the required maximum and minimum number of measurements $s$ and $r$. In the main loop of the algorithm, QEMI randomly selects a conditional statement or a nested structure of multiple conditional statements from the dead code patterns $N_p$ supported by the specified quantum programming language (line 9). Based on $N_p$, QEMI generates a complete quantum program $P$ that contains the corresponding dead code segments $D_c$ (line 10). For a given program $P$, by modifying or transforming $D_c$, QEMI can obtain a new quantum program $P'$ that preserves the same semantics as $P$ (line 11). For all executable configurations, both programs $P$ and $P'$ are executed (line 13). Based on the results of the two programs, we perform different actions: 

1) For simultaneous crashes with different errors or a single crash, QEMI will save these programs and label them as \textit{crash} (line 15). 

2) For a no-crash situation, QEMI will utilize the Hellinger-based early stop strategy (lines 17-25). In this process, we iteratively add additional measurement times to the target program pair ($P$, $P'$). Each new set of results ($R$, $R'$) is merged with the previous ones (line 23) and checked against violations of the $Hellinger()$ (line 19). If the total number of measurements exceeds the maximum times $s$, the early-stop process will terminate, and a final probability distribution result will be used to determine whether QSS exists a bug. If $Hellinger()$ rejects the hypothesis of identical distributions, QEMI will save these programs and label them as \textit{wrong} (line 26). 

3) In all other cases, QEMI will not take any action. After reaching the maximum number of executions, QEMI will output all identified potential buggy program pairs.

\begin{algorithm}[htb]
	\renewcommand{\algorithmicrequire}{\textbf{Input:}}
	\renewcommand{\algorithmicensure}{\textbf{Output:}}
	\caption{: Quantum EMI}
	\label{alg:alg1}
        \begin{scriptsize}
	\begin{algorithmic}[1]
        \REQUIRE $i\gets$ Quantum Language Platform;  $n \gets$ Number of qubits;  $\delta \gets$ Hellinger threshold \\
        \ENSURE $B\gets$ set of buggy programs
            \STATE $N \gets 2^n$
		\STATE $B \gets \emptyset$
            \STATE $config \gets ConfigChoice(i)$ 
		\STATE $G \gets ProgramGenerator(i)$ 
            \STATE $E \gets Executor(i)$
            \STATE $s \gets S(\delta, N)$  
            \STATE $r \gets S(\delta, N^{1/2})$  
            \FOR{$j\gets 0$ to $MaxIter$}
            \STATE $N_p \gets DeadCodePattern(i)$
            \STATE $P,\ D_c \gets G.generate(N_p)$
            \STATE $P' \gets Remove(P, D_c)$
            \FOR{$\sigma \in config$}
            \STATE $R,\ R' \gets E.execute(P, \  P',\  r, \  \sigma)$
            \IF{$Error(R) \neq Error(R')$}
            \STATE $B \gets B \ \cup \  \{(P,\ P',\ 'crash')\}$
            \ELSIF{$Error(R)=Error(R')=None$}
            \STATE $m_r \gets r$
            \WHILE{$m_r < s$}
            \IF{$Hellinger(Output(R,\ R'))$}
            \STATE $Break$
            \ENDIF 
            \STATE $m_r \gets m_r+r$
            \STATE $R,\ R' \gets Merge(E.execute(P, \  P',\  r, \  \sigma))$
            \ENDWHILE
            \IF{$\neg Hellinger(Output(R,\ R'))$} 
            \STATE $B \gets B \ \cup \  \{(P,\ P',\ 'wrong')\}$
            \ENDIF 
            \ENDIF 
            \ENDFOR
            \ENDFOR
            \RETURN $B$
	\end{algorithmic}

        \end{scriptsize}
\end{algorithm}

\begin{table}[t]
\scriptsize
\setlength{\tabcolsep}{4pt}
\centering
\caption{Real-world bugs found by QEMI.}
\label{tab:bugsum}
\tiny
{
\setlength{\parskip}{0pt}
\rowcolors{2}{gray!10}{white}

\begin{tabularx}{\columnwidth}{@{}c >{\raggedright\arraybackslash}p{1.0cm} c c >{\raggedright\arraybackslash}Y >{\raggedright\arraybackslash}Y@{}}
\toprule
ID & Report & Platform & Status & Crash Message & Bug Reason \\ \midrule

1  & \#14254 \#14338 \#14385 & Qiskit & Fixed
   & Unable to translate the operations in the circuit: [\textit{if\_else}, \textit{clifford}] to backend’s basis: \{h, mcphase, ...\}.
   & Failing to decompose a Clifford operator within a control-flow block. \\

2$^*$ & \#14521 & Qiskit & Fixed
    & No viable alternative at input ``\_''.
    & The \textit{ForLoopOp} does not support dynamic range. \\

3  & \#14407 & Qiskit & Fixed
   & Invalid param type \textit{complex} for gate \textit{initialize\_dg}.
   & Instruction \textit{Initialize} is not invertible. \\

4  & \#14635 \#14409 & Qiskit & Fixed
   & \textit{inverse()} not implemented for \textit{for\_loop}.
   & \textit{for\_loop} does not have inverse methods. \\

5  & \#14645 & Qiskit & Fixed
   & \textit{Clifford} object has no attribute \textit{is\_parameterized}.
   & Cliffords do not have \textit{is\_parameterized} methods. \\

6  & \#14319 & Qiskit & Confirmed
   & Program enters an infinite loop after \textit{RemoveFinalMeasurements} optimization.
   & -- \\

7  & \#14646 \#14647 & Qiskit & Confirmed
   & Called \textit{Option::unwrap()} on a \textit{None} value.
   & Failing to revalidate analysis before \textit{ConsolidateBlocks} after intermediate transformation. \\

8  & \#14522 & Qiskit & Reported
   & Declarations of type \texttt{int} are not supported.
   & -- \\ \midrule

9  & \#2534 & Q\# & Fixed
   & The generated program and its EMI variant results are in different distributions.
   & A bug in the gate queueing optimization of the underlying simulator. \\ \midrule

10 & \#7473 & Cirq & Fixed
   & Coefficient is not unitary.
   & The identity gate is not preserving the qubit order as expected. \\

11 & \#7472 & Cirq & Confirmed
   & Measurement key \textit{c} missing when testing classical control.
   & Incorrectly skips commutativity checks on disjoint qubits, missing classical control constraints. \\

12 & \#7474 & Cirq & Confirmed
   & Measurement key \textit{c} missing when testing classical control.
   & Failing to reverse the operation order within moments during circuit reversal. \\
\bottomrule
\end{tabularx}
}
\end{table}

\section{Experimental Evaluation}
\label{sec:result}
Our experiments focus on the following research questions (\textbf{RQ}s): 
\begin{itemize}
    \item \textbf{RQ1: }What kinds of real-world bugs did QEMI uncover?
    \item \textbf{RQ2: }How does QEMI compare with the baseline methods on bug detection? 
    \item \textbf{RQ3: }How much does QEMI’s early termination measurement strategy improve efficiency?
\end{itemize}

\subsection{Testing Setup}
\paragraph{Hardware and QSSes Version} We conducted our experiments on a workstation running Ubuntu 20.04, equipped with an Intel Core i9-10940X processor (14 cores, 28 threads) and 208 GB of RAM. We conduct our tests on the latest version of each QSS, including Qiskit (versions 2.0.0 to 2.1.0), Cirq (versions 1.6.0.dev20250702012506), and Q\# (version 1.17).
\paragraph{Metric for Behavior Checking} In our experiments, we use a Hellinger distance threshold of $\delta = 0.1$ to determine whether two output distributions differ. 

\subsection{RQ1: Bug Summary}

Table~\ref{tab:bugsum} summarizes the real-world bugs that QEMI found in three mainstream quantum software stacks—Qiskit, Cirq, and Q\#—by running each stack for 24 hours independently on its latest release. In total, we identified 12 bugs: 11 crash-inducing issues and 1 output-distribution discrepancy. Reports with the same or similar root cause were grouped under a single entry; one entry (marked with a star) is a duplicated issue that was independently rediscovered. For each entry, we list the platform, the report ID, status (Reported/Confirmed/Fixed), crash message, and a brief root-cause summary. As of submission, 7/12 have been fixed, 4 have been confirmed by upstream developers, and 1 is pending confirmation. The bugs span low-level failures (e.g., crashes in circuit transformation/optimization) and higher-level issues (e.g., semantic mismatches or incorrect output distributions). We next analyze their causes, focusing on quantum control flow and how dead code exposes these faults.

QEMI identified \emph{four} bugs (IDs 1, 2, 4, and 7) directly tied to quantum control flow, triggered either by optimizations applied to control structures or by transformations across quantum languages—highlighting that code transformation remains challenging for current QSSes in the presence of quantum control.

The remaining bugs are not directly caused by control flow, yet EMI was still crucial to reveal them. Dead code in the original program contributed in two ways. First, the faulty code itself lay inside the dead region; removing it in EMI variants made only the original crash (e.g., Qiskit \#14338, Cirq \#7472). Second, the dead region contained a control structure; some passes do not support control, so the original program silently skipped an optimization while the EMI variant proceeded along the unsupported path and crashed (e.g., Qiskit \#14254, \#14407).


\begin{tcolorbox}
\textbf{Answer to RQ1: }QEMI identified $12$ unique bugs in the latest versions of QSSes, including $1$ duplicate. Among them, $11$ have been confirmed or fixed by the official developers of the respective QSSes.  
\end{tcolorbox}

\subsection{RQ2: Baselines Comparison}

We compare QEMI with three baseline methods: QDiff, MorphQ, and QuteFuzz, which are automated techniques for testing QSSes. All experiments are conducted on the Qiskit platform, targeting versions 2.0.0 to 2.1.0. To ensure a fair comparison under consistent experimental conditions, we independently ran the complete workflow for each baseline method for 24 hours, covering program generation, execution, and checking for behavioral consistency.

\begin{table}[htb]
\centering
\captionsetup{skip=2pt} 
\small
{
\setlength{\tabcolsep}{3pt}      
\renewcommand{\arraystretch}{0.9}

\caption{Coverage after 24 hours of execution on each method.}
\label{tab:coverage}

\begin{tabular}{@{}p{1.5cm} p{1.5cm} p{1.5cm} p{1.5cm}@{}}
\toprule
QEMI & QDiff & MorphQ & QuteFuzz \\
\midrule
14.80\% & 6.24\% & 9.58\% & 14.18\% \\
\bottomrule
\end{tabular}
}
\end{table}

\subsubsection{Code Coverage}
Table~\ref{tab:coverage} presents the code coverage achieved by QEMI and three baseline methods. Among the four approaches, QEMI achieves the highest code coverage. This is likely because, unlike QDiff and MorphQ, which generate test programs using only basic gates, QEMI combines optimization passes and control flow structures during the test generation process. In this respect, QuteFuzz adopts a similar strategy by incorporating both optimization passes and control flow. However, QEMI has a slight advantage over QuteFuzz because it additionally invokes API functionalities from the QSS during generation, further contributing to its higher coverage.

\subsubsection{Bug-exposing capability}

\textcolor{black}{In our experiments, among the three baselines, only QuteFuzz triggered a bug, which corresponds to the known Qiskit issue \href{https://github.com/Qiskit/qiskit/issues/13165}{\#13165}. Under the same experimental settings, we did not observe any bugs triggered by QDiff or MorphQ. This suggests that QEMI is complementary to existing approaches and can trigger failure modes that are difficult to expose without EMI-style dead-code construction.}

\textcolor{black}{We conjecture that QEMI missed this issue due to its evaluation strategy. QEMI executes the original program and its EMI variant independently and reports a bug only when their observed behaviors diverge. If both executions crash in the same manner, QEMI treats them as behaviorally consistent and thus does not flag the failure. Consequently, crash bugs that manifest identically in both the original and EMI-augmented executions can be missed by this strategy.}

\begin{tcolorbox}
\textbf{Answer to RQ2: }Compared to prior work, QEMI can detect bugs that are difficult for existing approaches to expose. Although it performs fewer tests within the same time, the programs generated by QEMI achieve higher code coverage. 
\end{tcolorbox}

\subsection{RQ3: Effectiveness of the Early Stop Strategy}


We assess the effectiveness of QEMI’s early stopping by measuring its impact on checking efficiency and on the accuracy of distributional consistency validation. We quantitatively evaluate 2{,}000 Q\# programs with 6 and 8 qubits generated by QEMI, using a Hellinger threshold of $\delta=0.1$. Each program is measured up to $S_{\text{max}}$ shots while tracking whether its Hellinger distance ever exceeds the threshold to detect potential false negatives (i.e., classifying discrepant pairs as equivalent). Across all cases where early stopping was triggered, the Hellinger distance never exceeded the threshold even when measurements continued to $S_{\text{max}}$, indicating a low false-negative rate in practice. We also record the shot count at which each program terminates.

\begin{table}[htbp]
\centering
\caption{Distribution of Early Stop Shots ($\delta=0.1$)}
{\tiny                      
\setlength{\tabcolsep}{3pt}
\renewcommand{\arraystretch}{0.95}

\subcaption*{(a) Qubit number $n=6$}
\begin{tabularx}{\linewidth}{@{}l *{10}{>{\centering\arraybackslash}X}@{}}
\toprule
\makecell[l]{\textbf{multiple}\\($\times S_{\text{round}}$)} 
 & 2 & 3 & 4 & $S_{\text{std}}$ & 5 & 6 & 7 & 8 & 9 & $S_{\text{max}}$ \\
\midrule
\textbf{shots} & 952 & 1428 & 1904 & 2263 & 2380 & 2856 & 3332 & 3808 & 4284 & 4526 \\
\textbf{count} & 1806 & 80 & 47 & -- & 53 & 13 & 1 & 0 & 0 & -- \\
\bottomrule
\end{tabularx}

\subcaption*{(b) Qubit number $n=8$}
\begin{tabularx}{\linewidth}{@{}l *{15}{>{\centering\arraybackslash}X}@{}}
\toprule
\makecell[l]{\textbf{multiple}\\($\times S_{\text{round}}$)} 
 & 2 & 3 & 4 & 5 & 6 & 7 
 & \makecell[l]{$S_{\text{std}}$\\(8)} 
 & 9 & 10 & 11 & 12 & 13 & 14 & 15 
 & \makecell[l]{$S_{\text{max}}$\\(16)} \\  
\midrule
\textbf{shots} & 1600 & 2400 & 3200 & 4000 & 4800 & 5600 & 6400
               & 7200 & 8000 & 8800 & 9600 & 10400 & 11200 & 12000 & 12800 \\
\textbf{count} & 1832 & 78 & 44 & 10 & 6 & 5 & 0
               & 5 & 14 & 6 & 0 & 0 & 0 & 0 & 0 \\
\bottomrule
\end{tabularx}
} 
\label{tab:earlystop-both}
\end{table}
The detailed distributions are shown in Table~\ref{tab:earlystop-both}.
Each column reports the number of programs that terminated after performing a specific number of measurements, which is always a multiple of the round size $S_{\text{round}}$. The first row labels each column with its corresponding multiple of $S_{\text{round}}$ to clarify its relation to the standard and maximum budgets.
The column marked $S_{\text{std}}$ corresponds to the number of measurements required to achieve a $2/3$ confidence level based on theoretical bounds.  
We use $S_{\text{std}}$ as the baseline for the number of measurements needed to verify distributional consistency between program outputs.  
Accordingly, we quantify the speedup enabled by the early stopping strategy by computing the ratio between the total number of measurements actually performed and the baseline total, as follows:

\begin{equation}
\text{Speedup Ratio} = 1 - \frac{\sum_{i=1}^{T} S_i^{\text{early}}}{T \cdot S_{\text{std}}}
\end{equation}
where $S_i^{\text{early}}$ denotes the number of shots required for the $i$-th program under the early stopping strategy, and $T$ is the total number of programs.  
Using this formula, we observe a speedup of \textbf{53.83\%} for $n = 6$ and \textbf{72.21\%} for $n = 8$.

The columns to the right of the $S_{\text{std}}$ divider represent programs for which early stopping was not triggered within the standard budget. These programs continued sampling beyond $S_{\text{std}}$ but terminated before reaching the maximum limit $S_{\text{max}}$. Notably, all of these programs would be incorrectly classified as inequivalent under a fixed-budget strategy based solely on $S_{\text{std}}$ measurements, despite their distributions ultimately satisfying the Hellinger threshold. In contrast, QEMI’s early stopping strategy correctly identified these programs as equivalent. This highlights the advantage of QEMI's adaptive strategy in avoiding such false positives.


\begin{tcolorbox}
\textbf{Answer to RQ3: }QEMI’s early stopping strategy demonstrates significant acceleration in practical testing scenarios. It effectively reduces measurement cost while also mitigating false positives, thereby improving the overall efficiency and reliability of QSS testing. 
\end{tcolorbox}


\section{Threats to Validity}
\label{sec:threats}

Our implementation currently supports three major QSSes: Q\#, Qiskit, and Cirq. However, other frameworks such as Pytket also support quantum circuits with control flow. Extending QEMI to additional QSSes would require implementing corresponding quantum random program generators and constructing appropriate dead-code templates for the target platforms. Nonetheless, the core idea of testing QSSes by generating semantically equivalent quantum programs via EMI remains generalizable.

The early stopping strategy that we employ to accelerate the validation of behavioral consistency is primarily based on empirical observations rather than formal theoretical guarantees. Although no false positives or false negatives have been observed in our experiments so far, this outcome is not theoretically guaranteed. Nevertheless, our strategy enables testing more program instances within the same time budget, which we consider a worthwhile trade-off.

Finally, our evaluation is currently limited to simulations due to classical hardware limitations. All experiments are conducted on simulators, restricting us to circuits with up to $8$ qubits. While we have not yet performed large-scale evaluations on real quantum hardware, we believe our approach is inherently applicable to real-device scenarios, as it relies solely on observable measurement distributions and does not depend on backend-specific simulation properties. Nevertheless, the presence of hardware noise and device-specific behavior may introduce new challenges, which we leave for future investigation.

\section{Related Work}
\label{sec:related work}

Our methodology is closely connected to prior research in classical compiler testing as well as the testing of quantum software stacks. In this section, we review relevant and representative studies organized into two subsections. Section~\ref{subsec:cct} discusses classical compiler testing, and Section~\ref{subsec:qsst} focuses on testing methodologies for quantum software stacks.

\subsection{Classical Compiler Testing. }
\label{subsec:cct}

Classical compiler testing has been extensively studied in the programming languages and systems community.
\textit{Csmith}~\cite{csmith} is a widely used random program generator based on differential testing that detects crashing or miscompilation bugs by generating random C programs and comparing their behavior across compilers.
\textit{Yarpgen}~\cite{yarpgen2018}, developed at Intel, further improves random generation by statically tracking constraints to avoid undefined behavior and by steering toward optimization-relevant patterns.
\textit{Orion}~\cite{emi} was the first framework to apply Equivalence Modulo Inputs (EMI) to C compilers, and subsequent tools such as \textit{CLsmith}~\cite{clsmith} extended similar ideas to other languages.
These approaches have been highly effective for exposing compiler bugs in the classical setting.
However, quantum programs bring domain-specific challenges—superposition, entanglement, and measurement-induced randomness—and operate on qubits rather than classical bits, making a direct transfer of classical techniques to quantum software stacks non-trivial.

\subsection{Quantum Software Stacks Testing. }
\label{subsec:qsst}

Several studies have focused on testing QSSes. \textit{QDiff}~\cite{qdiff} detects inconsistencies by applying semantics-preserving transformations, such as equivalent gate replacements and optimization level changes, to a small set of hand-written quantum programs and checks whether their behaviors remain consistent. \textit{MorphQ}~\cite{morphq} adopts a metamorphic testing approach by applying a broader range of transformations to randomly generated quantum programs. Then it verifies whether the transformed variants exhibit the same behavior as the original program. Neither QDiff nor MorphQ addresses testing of quantum control flow, and our approach does not overlap with their transformation techniques, making them complementary to our method. Among existing tools, \textit{QuteFuzz}~\cite{qutefuzz} is the only one that explicitly targets quantum control flow. It generates random quantum programs with control constructs and examines behavioral consistency under varying optimization levels or injected passes. However, QuteFuzz does not incorporate EMI techniques, which are central to our design. Moreover, QuteFuzz and MorphQ do not incorporate certain built-in circuit structures—such as QFT~\cite{Nielsen_Chuang_2010}—that are directly available in QSS APIs when generating random programs. In contrast, our approach addresses this limitation by explicitly supporting such predefined modules. \textit{Upbeat}~\cite{upbeat} focuses specifically on the Q\# platform. Instead of testing for semantic equivalence, it extracts constraints from the use of Q\# APIs and checks whether the compiler respects these constraints, thereby detecting potential boundary bugs. While effective in exposing API-level violations, Upbeat does not target the behavioral consistency of compiled quantum programs or address quantum control-flow structures.

\section{Conclusion}
\label{sec:conclusion}

We presented QEMI, the first framework that adapts the classical Equivalence Modulo Inputs (EMI) approach to quantum software testing. QEMI automatically generates quantum programs with dead code and corresponding EMI variants with that code removed; by executing both and checking for behavioral inconsistencies, it reveals defects in quantum software platforms.
In our evaluation, QEMI uncovered 12 unique bugs across three major QSSes (Qiskit, Q\#, Cirq), 11 of which have been confirmed or fixed. Beyond detection, an adaptive early-stopping strategy reduces measurement cost while preserving equivalence-checking accuracy and avoiding false positives that arise under fixed-budget testing, improving the robustness and scalability of QSS testing.

For future work, pattern-based insertion of dead code can be replaced with analyses that detect it automatically, either statically (via control and data-flow reasoning across quantum–classical boundaries) or dynamically when feasible, improving scalability and enabling reuse by QSS optimizers. Beyond simple dead-code removal, richer semantics-preserving transformations within the identified regions could be applied; such mutations are expected to increase the likelihood of surfacing subtle defects while maintaining equivalence for the target inputs.

\begin{credits}
\subsubsection{\ackname} This work was supported in part by the JST SPRING Grant No.\ JPMJSP2136, and JSPS KAKENHI grants Grant No.\ JP23K28062, No.\ JP24K02920, and No.\ JP24K14908. 

\end{credits}

%
%
%
\bibliographystyle{splncs04}
\bibliography{ref}
%




\end{document}